\documentclass[MSNbibl,nameyear,dvips]{arxstspdf}
\usepackage{graphicx}
\usepackage{flushend}
\usepackage{stfloats}

\volume{27}
\issue{1}
\pubyear{2012}
\firstpage{82}
\lastpage{94}
\doi{10.1214/11-STS383}

\makeatletter
\newtheorem{theorem}{Theorem}
\newtheorem{lemma}{Lemma}
\newtheorem{coro}{Corollary}
\newcommand{\C}{|}
\newcommand{\Y}{\tilde{Y}}
\def\hp{\hat{p}}
\newcommand{\fraca}[2]{{#1}/{#2}}
\makeatother

\begin{document}
\begin{frontmatter}
\vspace*{6pt}
\title{From Minimax Shrinkage Estimation to~Minimax Shrinkage Prediction}
\runtitle{Minimax Shrinkage Prediction}

\begin{aug}
\author[a]{\fnms{Edward I.} \snm{George}\corref{}\ead[label=e1]{edgeorge@wharton.upenn.edu}},
\author[b]{\fnms{Feng} \snm{Liang}\ead[label=e2]{liangf@illinois.edu}}
\and
\author[c]{\fnms{Xinyi} \snm{Xu}\ead[label=e3]{xinyi@stat.osu.edu}}
\runauthor{E. I. George, F. Liang and X. Xu}

\affiliation{The Wharton School, University of Illinois at
Urbana-Champaign and  Ohio State University}

\address[a]{Edward I. George is Professor, Department of Statistics,
The Wharton School, Philadelphia, PA
19104-6340, USA
\printead{e1}.}
\address[b]{Feng Liang is Associate Professor, Department of
Statistics, University of Illinois at Urbana-Champaign, Champaign,
IL 61820, USA \printead{e2}.}
\address[c]{Xinyi Xu is Assistant Professor, Department of Statistics,
The Ohio State University, Columbus, OH 43210-1247, USA \printead{e3}.}

\end{aug}

%
\begin{abstract}
In a remarkable series of papers beginning in 1956, Charles Stein set
the stage for the future development of minimax shrinkage estimators of
a multivariate normal mean under quadratic loss. More recently,
parallel developments have seen the emergence of minimax shrinkage
estimators of multivariate normal predictive densities under
Kullback--Leibler risk. We here describe these parallels emphasizing
the focus on Bayes procedures and the derivation of the superharmonic
conditions for minimaxity as well as further developments of new
minimax shrinkage predictive density estimators including multiple
shrinkage estimators, empirical Bayes estimators, normal linear model
regression estimators and nonparametric regression estimators.
\end{abstract}

%

\end{frontmatter}

\section{The Beginning of the Hunt for Minimax Shrinkage
Estimators}\label{sec1}
Perhaps the most basic estimation problem in Sta\-tistics is the
canonical problem of estimating a multivariate normal mean. Based on
the observation of a~$p$-dimensional multivariate normal random \mbox{variable}
%
\begin{equation}
X \C\mu\sim N_p(\mu, I),
\end{equation}
the problem is to find a suitable estimator $\hat\mu(x)$ of~$\mu$.
The celebrated result of Stein
(\citeyear{Ste56}) dethroned $\hat\mu_{\mathrm{MLE}}(x) =
x$, the\vadjust{\goodbreak} maximum likelihood and best location invariant estimator for
this problem, by showing that, when $p \ge3$, $\hat\mu_{\mathrm{MLE}}$ is
inadmissible under quadratic loss
%
\begin{equation}
R_Q(\mu, \hat\mu) = E_\mu\|\hat\mu(X) - \mu\|^2.
\end{equation}
From a decision theory point of view, an important part of the appeal
of $\hat\mu_{\mathrm{MLE}}$ was the protection offered by its minimax
property. The worst possible risk~$R_Q$ incurred by $\hat\mu_{\mathrm{MLE}}$
was no worse than the worst possible risk of any other estimator.
Stein's result implied the existence of even better estimators that
offered the same minimax protection. He had begun the hunt for these
better minimax estimators.

In a remarkable series of follow-up papers Stein proceeded to set the
stage for this hunt. James and Stein (\citeyear{J61}) proposed a new closed-form
minimax shrinkage estimator
%
\begin{equation}\label{JSestimator}
\hat\mu_{\mathrm{JS}}(x) = \biggl (1- \frac{p-2}{\|x\|^2} \biggr) x,
\end{equation}
the now well-known James--Stein estimator, and
showed explicitly that
its risk was less than $R_Q(\mu,\allowbreak  \hat\mu_{\mathrm{MLE}}) \equiv p$ for every
value of $\mu$ when $p \ge3$, that is, it uniformly dominated $\hat
\mu_{\mathrm{MLE}}$. The appeal of $\hat\mu_{\mathrm{JS}}$ under~$R_Q$ was compelling.
It offered the same guaranteed minimax protection as $\hat\mu_{\mathrm{MLE}}$
while also offering the possibility of doing much better.

Stein
(\citeyear{Ste62N1}), though primarily concerned with improved confidence
regions, described a parametric empirical Bayes motivation for (\ref
{JSestimator}), describing\break how~$\hat\mu_{\mathrm{JS}}(x)$ could be seen as a
data-based approximation to the posterior mean
%
\begin{equation}\label{conjugate-coef}
E_\pi(\mu\C x) =  \biggl(1- \frac{1}{1+\nu} \biggr) x,
\end{equation}
the Bayes rule which minimizes the average risk $E_\pi R_Q(\mu, \hat
\mu)$ when $\mu\sim N_p(0, \nu I)$. He here also proposed the
positive-part James--Stein estimator\break $\hat\mu_{\mathrm{JS}+} = \max\{0, \hat
\mu_{\mathrm{JS}}\}$, a dominating improvement\break over~$\hat\mu_{\mathrm{JS}}(x)$, and
commented that ``it would be even better to use the Bayes estimate with
respect to a~reasonable prior distribution.'' These observations ser\-ved
as a clear indication that the Bayesian para\-digm was to play a major
role in the hunt for these new shrinkage estimators, opening up a new
direction that was to be ultimately successful for establishing large
new classes of shrinkage estimators.

Dominating fully Bayes shrinkage estimators soon emerged. Strawderman
(\citeyear{Str71}) proposed $\hat\mu_a(x) = \break E_{\pi_a}(\mu\C x)$, a class of
Bayes shrinkage estimators obtai\-ned as posterior means under priors
$\pi_a(\mu)$ for~which
%
\begin{equation}\label{pi_a}
\mu| s \sim N_p  (0, s  I  ),
\quad s \sim(1 + s)^{a-2}.
\end{equation}
Strawderman explicitly showed that $\hat\mu_a$ uniformly dominated
$\hat\mu_{\mathrm{MLE}}$ and was proper Bayes, when $p = 5$ and $a \in[ 0.5,1)$
or when $p \ge6$ and $a \in[0,1)$. This was especially interesting
because any proper Bayes was necessarily admissible and so could not be
improved upon.

Then, Stein
(\citeyear{Ste74,Ste81}) showed that $\hat\mu_H(x)$, the Bayes
estimator under the harmonic prior
%
\begin{equation}\label{pi_H}
\pi_{H}(\mu)= E_{\pi_{H}}(\mu\C x)= \|\mu\|^{-(p-2)},
\end{equation}
dominated $\hat\mu_{\mathrm{MLE}}$ when $p \ge3$. A special case of $\hat\mu
_a$ when $ a=2$, $\hat\mu_H$ was only formal Bayes because $\pi
_{H}(\mu)$ is improper. Undeterred, Stein pointed out that the
admissibility of $\hat\mu_H$ followed immediately from the general
conditions for the admissibility of generalized Bayes estimators laid
out by Brown
(\citeyear{Bro71}). A~further key element of the story was Brown's
(\citeyear{Bro71}) powerful result that all such generalized Bayes rules (including
the proper ones of course) constituted a~complete class for the problem
of estimating multivariate normal mean under quadratic loss. It was now
clear that the hunt for new minimax shrinkage estimators was to focus
on procedures with at least some Bayesian motivation.

Perhaps even more impressive than the fact\break that~$\hat\mu_H$ dominated
$\hat\mu_{\mathrm{MLE}}$ was the way Stein proved it. Making further use of
the rich results in Brown
(\citeyear{Bro71}), the key to his proof was the fact
that any posterior mean Bayes estimator under a prior $\pi(\mu)$ can
be expressed as
%
\begin{equation}\label{keyrep1}
\hat\mu_\pi(x) = E_\pi(\mu\C x) = x + \nabla\log m_\pi(x),
\end{equation}
where
%
\begin{equation}\label{margx}
m_\pi(x) \propto \int e^{-(x-\mu)^2/2}   \pi(\mu)\,d\mu
\end{equation}
is the marginal distribution of $X$ under $\pi(\mu)$.
[Here $\nabla= (\frac{\partial}{\partial x_1},\ldots, \frac
{\partial}{\partial x_p})'$ is the familiar gradient.]

At first glance it would appear that (\ref{keyrep1}) has little to do
with the risk. However, Stein noted that insertion of (\ref{keyrep1})
into $R_Q$, followed by expansion and an integration-by-parts identity,
now known as one of Stein's Lemmas, yields the following general
expression for the difference between the risks of $\hat\mu_\pi$ and~$\hat\mu_{\mathrm{MLE}}$:
%
\begin{eqnarray}\label{UBER1a}
&&R_Q(\mu, \hat\mu_{\mathrm{MLE}}) - R_Q(\mu, \hat\mu_\pi) \nonumber
\\[-8.5pt]
\\[-8.5pt]
&& \quad = E_\mu
\biggl[\|\nabla\log m_\pi(X)\|^2 - 2 \frac{\nabla^2 m_\pi(X)}{m_\pi
(X)} \biggr] \nonumber\\[-0.5pt] \label{UBER1b}
&& \quad  =  E_\mu \bigl[-4\nabla^2 \sqrt{m_\pi(X)}\big/\sqrt{m_\pi
(X)} \bigr].
\end{eqnarray}
(Here $\nabla^2 = \sum_i \frac{\partial^2}{\partial x_i^2}$ is the
familiar Laplacian.)

Because the bracketed terms in (\ref{UBER1a}) and (\ref{UBER1b}) do
not depend on $\mu$ (they are unbiased estimators of the risk
difference), the domination of $\hat\mu_{\mathrm{MLE}}$ by $\hat\mu_\pi$
would follow whenever $m_\pi$ was such that these bracketed terms were
nonnegative. As Stein noted, this would be the case in (\ref{UBER1a})
whenever $m_\pi$ was superhar\-monic,
$\nabla^2 m_\pi(x) \le0$, and in (\ref{UBER1b}) whenever $\sqrt
{m_\pi}$~was superharmonic,
$\nabla^2 \sqrt{m_\pi(x)} \le0$,
a weaker condition.

The domination of $\hat\mu_{\mathrm{MLE}}$ by $\hat\mu_H$ was seen now to be
attributable directly to the fact that the margi\-nal~(\ref{margx})
under $\pi_{H}$, a mixture of harmonic functions, is superharmonic
when $p \ge3$. However, such an explanation would not work for the
domination of $\hat\mu_{\mathrm{MLE}}$ by~$\hat\mu_a$, because the marginal
(\ref{margx}) under~$\pi_a$ in (\ref{pi_a}) is not superharmonic for
any $a < 1$. Indeed, as was shown later by
Fourdrinier,
Strawderman and Wells (\citeyear{FouStrWel98}), a superharmonic marginal cannot be obtai\-ned with any
proper prior. More importantly, however, they were able to establish
that the domination by $\hat\mu_a$ was attributable to the
superharmonicity of $\sqrt{m_{\pi_a}}$ under $\pi_a$ when $p \ge5$
(and Strawderman's conditions on $a$). In fact, it also followed from
their results that $\sqrt{m_{\pi_a}}$ is superharmonic when $a \in
[1,2)$ and $p \ge3$, further broadening the class of minimax improper
Bayes estimators.

Prior to the appearance of (\ref{UBER1a}) and (\ref{UBER1b}),
minimaxity proofs, though ingenious, had all been tailored to suit the
specific estimators at hand. The sheer generality of this new approach
was daunting in its scope. By restricting attention to priors that gave
rise to marginal distributions with particular properties, the minimax
properties of the implied Bayes rules would be guaranteed.

\section{The Parallels in the Predictive Estimation Problem Emerge}\label{sec2}
\label{sec:pred-emerges}

The seminal work of Stein concerned the canonical problem of how to
estimate $\mu$ based on an observation of $X \C\mu\sim N_p(\mu,
I)$. A more ambitious problem is how to use such an $X$ to estimate the
entire probability distribution of a future $Y$ from a normal
distribution with this same unknown mean $\mu$, the so-called
predictive density of $Y$. Such a predictive density offers a complete
description of predictive uncertainty.

To conveniently treat the possibility of different variances for $X$
and $Y$, we formulate the predictive problem as follows. Suppose $X \C
\mu\sim N_p(\mu, v_x I)$ and $Y \C\mu\sim N_p(\mu, v_y I)$ are
independent $p$-dimensional multivariate normal vectors with common
unknown\break mean $\mu$ but known variances $v_x$ and $v_y$. Letting $p(y
\C \mu)$ denote the density of $Y$, the problem is to find an
estimator $\hat p(y \C x)$ of $p(y \C \mu)$ based on the observation
of $X = x$ only. Such a problem arises naturally, for example, for
predicting $Y \C\mu\sim N_p(\mu, \sigma^2 I)$ based on the
observation of $X_1,\ldots, X_n \C\mu \mbox{ i.i.d.} \sim
N_p(\mu,\break
\sigma^2 I)$ which is equivalent to observing $\bar X \C\mu\sim
N_p(\mu,\break (\sigma^2/n) I)$. This is exactly our formulation with $v_x
= \sigma^2/n$ and $v_y = \sigma^2$.

For the evaluation of $\hat p(y \C x)$ as an estimator of $p(y \C \mu
)$, the analogue of quadratic risk $R_Q$ for the mean estimation
problem is the Kullback--Leibler (KL) risk
%
\begin{equation}\label{eq:risk}
R_{\mathrm{KL}}(\mu, \hat p) = \int p(x \C \mu) L(\mu, \hat p(\cdot \C x))\,dx,
\end{equation}
where $p(x \C \mu)$ denotes the density of $X$, and
%
\begin{equation}\label{eq:loss}
L( \mu, \hat p(\cdot\C x)) = \int p(y \C \mu) \log\frac{p(y \C
\mu)}{\hat p(y\C x)}\,dy
\end{equation}
is the familiar KL loss.

For a (possibly improper) prior distribution $\pi$ on~$\mu$, the
average risk\vadjust{\goodbreak} $r(\pi, \hat p) = \int R_{\mathrm{KL}}(\mu, \hat p) \pi(\mu)\,d\mu$ is minimized by the Bayes rule
%
\begin{eqnarray}\label{eq:Bayes}
\hat{p}_\pi(y \C x)&=& E_\pi[p(y \C \mu) \C x]\nonumber
\\[-8pt]
\\[-8pt]
&=& \int p(y \C \mu) \pi(\mu\C x)\,d\mu,
\nonumber
\end{eqnarray}
the posterior mean of $p(y \C \mu)$ under $\pi$ (Aitchison,
\citeyear{Ait75}). It follows from (\ref{eq:Bayes}) that $\hat
{p}_\pi(y \C x)$ is a proper probability distribution over $y$
whenever the marginal density of $x$ is finite for all $z$ (integrate
w.r.t. $y$ and switch the order of integration). Furthermore, the mean
of $\hat{p}_\pi(y \C x)$ (when it exists) is equal to $E_\pi( \mu\C
x)$, the Bayes rule for estimating $\mu$ under quadratic loss, namely
the posterior mean of $\mu$. Thus, $\hat{p}_\pi$ also carries the
necessary information for that estimation problem. Note also that
unless $\pi$ is a trivial point prior, such $\hat{p}_\pi(y \C x)$
will not be of the form of $p(y \C \mu)$ for any $\mu$. The range of
the Bayes rules here falls outside the target space of the densities
which are being estimated.

A tempting initial approach to this predictive density estimation
problem is to use the simple plug-in estimator $\hat{p}_{\mathrm{MLE}} \equiv
p(y \C\mu= \hat\mu_{\mathrm{MLE}})$ to estimate $p(y \C \mu)$, the
so-called estimative approach. This was the conventional wisdom until
the appearance of Aitchison
(\citeyear{Ait75}).
He showed that the plug-in estimator $\hat{p}_{\mathrm{MLE}}$ is uniformly
dominated under $R_{\mathrm{KL}}$ by
%
\begin{eqnarray}\label{eq:piU}
   \hat{p}_U(y \C x) &\equiv& E_{\pi_U}[p(y \C \mu) \C x]\hspace*{-17pt}\nonumber
\\[-7pt]
\\[-8pt]    &=& \frac{1}{\{
2\pi(v_x + v_y)\}^{\fraca{p}{2}}}   \exp \biggl\{ -\frac{\|y-x\|
^2}{2(v_x + v_y)}  \biggr\},\hspace*{-17pt}
\nonumber
\end{eqnarray}
the posterior mean of $p(y \C \mu)$ with respect to the uniform prior
$\pi_U(\mu)=1$, the so-called predictive approach. In a related vein,
Akaike (\citeyear{A78}) pointed out that, by Jensen's inequality, the Bayes rule
$\hat{p}_\pi(y \C x)$ would dominate the random plug-in estimator\break
$\hat{p}(y \C\mu= \hat\mu)$ when $\hat\mu$ is a random draw from
$\pi$. Strategies for averaging over $\mu$ were looking better than
plug-in strategies. The hunt for predictive shrinkage estimators had
turned to Bayes procedures.

Distinct from $\hat{p}_{\mathrm{MLE}}$, $\hat{p}_U$ was soon shown to be the
best location invariant predictive density estimator; see Murray
(\citeyear{Mur77})
and Ng
(\citeyear{Ng80}). That $\hat{p}_U$ is best invariant and minimax also
follows from the more recent general results of Liang and Barron (\citeyear{LiaBar04}), who also showed that $\hat{p}_U$ is admissible when $p = 1$.
The minimaxity of $\hat{p}_U$ was also shown directly by George,
Liang and Xu (\citeyear{GeoLiaXu06}). Thus, $\hat{p}_U$, rather than $\hat{p}_{\mathrm{MLE}}$, here
plays the role played by $\hat\mu_{\mathrm{MLE}}$ in the mean estimation
context. Not surprisingly, $\hat\mu_U = x$, the posterior mean under
the uniform prior $\pi_U$ is identical to $\hat\mu_{\mathrm{MLE}}$ in that context.

The parallels between the mean estimation problem and the predictive
estimation problem came into sharp focus with the stunning breakthrough
result of Komaki
(\citeyear{Kom01}). He proved that when $p \ge3$, $\hat{p}_U(y
\C x)$ itself is dominated by the Bayes rule
%
\begin{equation} \label{eq:pH}
\hat{p}_H(y \C x) = E_{\pi_H}[p(y \C \mu) \C x],
\end{equation}
under the harmonic prior $\pi_H(\mu)$ in (\ref{pi_H}) used by Stein
(\citeyear{Ste74}). Shortly thereafter Liang
(\citeyear{Lia02}) showed\break that~$\hat{p}_U(y \C
x)$ is dominated by the proper Bayes ru\-le~$p_a(y \C  x)$ under~$\pi
_a(\mu)$ for which
%
\begin{equation}\label{eq:pa}
\mu| s \sim N_p  ( 0, s  v_0 I  ),
\quad s \sim(1 + s)^{a-2}, \label{sprior}
\end{equation}
when $v_x \le v_0$, and when $p = 5$ and $a \in[0.5,1)$ or \mbox{$p \ge6$}
and $a \in[0,1)$, the same conditions that Strawderman had obtained
for his estimator. Note\break that~$\pi_a(\mu)$ in (\ref{eq:pa}) is an
extension of (\ref{pi_a}) which depends on the constant $v_0$. As
before, $\pi_H(\mu)$ is the special case of~$\pi_a(\mu)$ when $a
=2$. Note that $\hat{p}_U$ is now playing the ``straw-man'' role that
was played by $\hat\mu_{\mathrm{MLE}}$ in the mean estimation problem.

\section{A Unified Theory for Minimax Predictive Density
Estimation}\label{sec3}\label{sec:theory}

The proofs of the domination of $\hat{p}_U$ by $\hat{p}_H$ in Komaki
(\citeyear{Kom01}) and by $\hat{p}_a$ in
Liang
(\citeyear{Lia02}) were both tailored to the specific forms of the dominating
estimators. They did not make direct use of the properties of the
induced marginal distributions of $X$ and~$Y$. From the theory
developed by Brown
(\citeyear{Bro71}) and Stein
(\citeyear{Ste74}) for the mean estimation
problem, it was natural to ask if there was a theory analogous to (\ref
{keyrep1})--(\ref{UBER1b}) which would similarly unify the domination
results in the predictive density estimation problem.

As it turned out, just such a theory was established in George,
Liang and Xu (\citeyear{GeoLiaXu06}), the main results of which we now proceed to describe.
The story begins with a representation, analogous to Brown's
representation $\hat\mu_\pi(X) = E_\pi(\mu\C X) = X + \nabla\log
m_\pi(X)$ in (\ref{keyrep1}), that is available for posterior mean
Bayes rules in the predictive density estimation problem. A key element
of the representation is the form of the marginal distributions for our
context which we denote by
%
\begin{equation}\label{eq:margz}
m_\pi(z;v) = \int p(z \C \mu) \pi(\mu)\,d\mu
\end{equation}
for $Z \C\mu\sim N_p(\mu,vI)$ and a prior $\pi(\mu)$. In terms of
our previous notation (\ref{margx}), $m_\pi(z) = m_\pi(z;1)$.

%
\begin{lemma}\label{thm:pform} The Bayes rule $\hat{p}_\pi(y \C x)$
in (\ref{eq:Bayes}) can be expressed as
%
\begin{equation}\label{eq:mform}
\hat{p}_\pi(y \C x) =\frac{m_\pi(w; v_w)} {m_\pi(x ; v_x)}  \hat
{p}_U(y \C x),
\end{equation}
where $\hat{p}_U(y \C x)$ is the Bayes rule under $\pi_U(\mu) = 1$
given by (\ref{eq:piU}), $m_\pi(x ; v_x)$ is the marginal
distribution of $X$, and $m_\pi(w; v_w)$, where $v_w = \frac{v_x
v_y}{v_x + v_y}$, is the marginal distribution of $W = \frac{ v_y X +
v_x Y} {v_x + v_y}$ for independent $X \C\mu\sim N_p(\mu, v_x I)$
and $Y \C\mu\sim N_p(\mu, v_y I)$.
\end{lemma}

Lemma \ref{thm:pform} shows how the form of $\hat{p}_\pi(y \C x)$ is
determined entirely by $\hat{p}_U(y \C x)$ and the form of $m_\pi
(x;v_x)$ and $m_\pi(w; v_w)$. The essential step in its derivation is
to factor the joint distribution of $x$ and $y$ into terms including a
function of the sufficient statistic $w$. Inserting the representation
(\ref{eq:mform}) into the risk~$R_{\mathrm{KL}}$ leads immediately to the
following unbiased estimate for the $\mathrm{KL}$ risk difference between $\hat
{p}_U(y \C x)$ and~$\hat{p}_\pi(y \C x)$:
%
\begin{eqnarray}\label{eq:uber3}
&& \quad R_{\mathrm{KL}}(\mu, \hat{p}_U) - R_{\mathrm{KL}}(\mu,
\hat{p}_\pi)\nonumber\\
&&  \quad \quad  =  \int\int p(x \C \mu)  p(y \C \mu) \log\frac{\hat{p}_\pi
(y \C x)}{\hat{p}_U(y\C x)} \,dx\,dy \\
&& \quad  \quad  =  E_{\mu,v_w} \log m_\pi(W; v_w) - E_{\mu, v_x} \log m_\pi(X ;
v_x).
\nonumber
\end{eqnarray}

As one can see from (\ref{eq:uber3}) and the fact that $v_w = \frac
{v_x v_y}{v_x + v_y} < v_x$, $\hat{p}_U(y \C x)$ would be uniformly
dominated by $\hat{p}_\pi(y \C x)$ whenever $E_{\mu,v} \log m_\pi
(Z; v)$ is decreasing in $v$. As if by magic, the sign of $\frac
{\partial}{\partial v} E_{\mu,v} \log m_\pi(Z;v)$ turned out to be
directly linked to the same unbiased risk difference estimates (\ref
{UBER1a}) and (\ref{UBER1b}) of Stein
(\citeyear{Ste74}).

%
\begin{lemma}\label{thm:rderiv}
%
\begin{eqnarray}\label{uber2a}
&& \quad \frac{\partial}{\partial v} E_{\mu,v} \log m_\pi(Z;v)\nonumber
\\[-8pt]
\\[-8pt]
&&  \quad \quad  =  E_{\mu,v}  \biggl[\frac{ \nabla^2 m_\pi(Z ;v)}{m_\pi(Z ;v)}
-
\frac{1}{2} \|\nabla\log m_\pi(Z ;v)\|^2  \biggr]
\nonumber \\
\label{uber2a}
&&  \quad \quad  =  E_{\mu,v} \bigl [2\nabla^2 \sqrt{m_\pi(Z ;v)}\big/\sqrt{m_\pi(Z
;v)} \bigr].
\end{eqnarray}
\end{lemma}

The proof of Lemma \ref{thm:rderiv} relies on
Brown's representation, Stein's Lemma, and the fact that any normal
marginal distribution $m_\pi(z; v)$ satisfies
%
\begin{equation}
\frac{\partial}{\partial v} m_\pi(z;v) = \frac{1}{2}\nabla^2 m_\pi(z;v),
\end{equation}
the well-known heat equation which has a long history in science and
engineering; for example, see Steele
(\citeyear{Ste01}). Combining (\ref
{eq:uber3}) and Lemma \ref{thm:rderiv} with the fact that $\hat
{p}_U(y \C x)$ is minimax yields the following general conditions for
the minimaxity of a predictive density estimator, conditions analogous
to those obtained by Stein for the minimaxity of a normal mean estimator.

%
\begin{theorem}\label{theo1} If $m_\pi(z ; v)$ is finite for all $z$, then $\hat
{p}_\pi(y \C x)$ will be minimax if either of the following hold for
all $v_w \le v \le v_x$:
\begin{longlist}[(ii)]
\item[(i)]$m_\pi(z;v)$ is superharmonic.
\item[(ii)]$\sqrt{m_\pi(z;v)}$ is superharmonic.
\end{longlist}
\end{theorem}

Although condition (i) implies the weaker condition~(ii) above, it is
included because of its convenience when it is available. Since a
superharmonic prior always yields a superharmonic $m_\pi(z;v)$ for all~$v$, the following corollary is immediate.

%
\begin{coro}\label{cor1} If $m_\pi(z ; v)$ is finite for all $z$,~then $\hat
{p}_\pi(y \C x)$ will be minimax if $\pi(\mu)$ is superharmonic.
\end{coro}

Because $\pi_H$ is superharmonic, it is immediate from Corollary~\ref{cor1}
that $\hat{p}_H$
is minimax. Because $\sqrt{m_a(z ; v)}$ is superharmonic for all $v$
(under suitable conditions on $a$), it is immediate from Theorem~\ref{theo1} that~$\hat{p}_a$ is~mi\-nimax. It similarly follows that any of the improper
superharmonic $t$-priors of Faith
(\citeyear{Fai78}) or any of the proper
generalized $t$-priors of Fourdrinier,
Strawderman and Wells (\citeyear{FouStrWel98})
yield minimax Bayes rules.

\begin{figure*}

\includegraphics{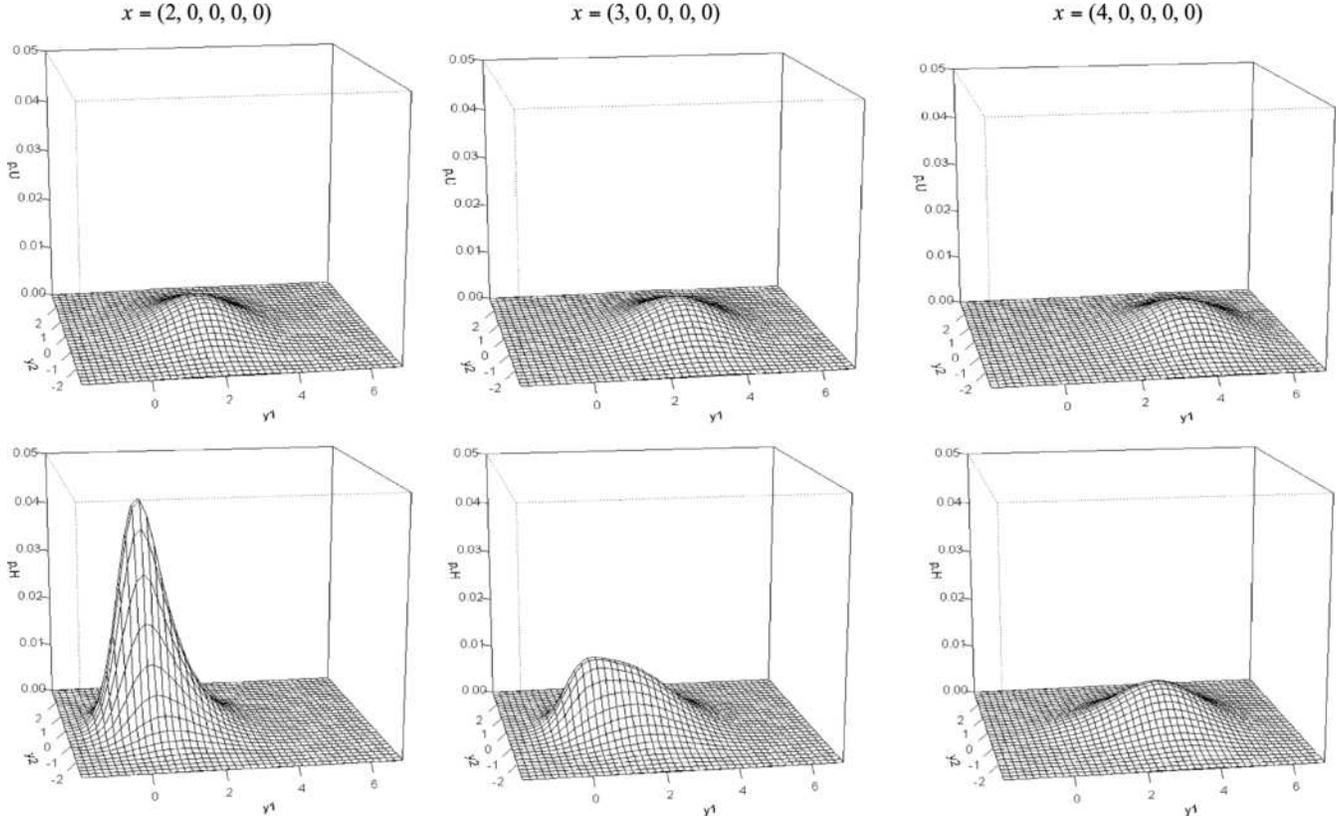}

\caption{Shrinkage of $\hp_U(y \C x)$ to obtain $\hp_H(y \C x)$ when $v_x = 1$,
$v_y = 0.2$ and $p=5$.  Here $y = (y_1, y_2, 0, 0, 0)'$.}
\label{fig1}\end{figure*}

The connections between the unbiased risk difference estimates for the
KL risk and quadratic risk problems ultimately yields the following identity:
%
\begin{eqnarray}
&& \quad R_{\mathrm{KL}}(\mu, \hat{p}_U) - R_{\mathrm{KL}} (\mu, \hat{p}_\pi)\nonumber
\\[-8pt]
\\[-8pt]
&&  \quad \quad= \frac{1}{2} \int_{v_w}^{v_x} \frac{1}{v^2} [ R_Q(\mu, \hat\mu
_U) - R_Q(\mu, \hat\mu_\pi)  ]_v \,dv,
\nonumber
\end{eqnarray}
explaining the parallel minimax conditions in both problems. Brown,
George and Xu (\citeyear{BroGeoXu08}) used this identity to further draw out connections
to establish sufficient conditions for the admissibility of Bayes rules
under KL loss, conditions analogous to those of Brown
(\citeyear{Bro71}) and Brown
and Hwang (\citeyear{BroHwa82}), and to show that all admissible procedures for the KL
risk problems are Bayes rules, a direct parallel of the complete class
theorem of Brown
(\citeyear{Bro71}) for quadra\-tic risk.

\section{The Nature of Shrinkage in Predictive Density Estimation}\label{sec4}

The James--Stein estimator $\hat\mu_{\mathrm{JS}}(x)$ in (\ref{JSestimator})
provided an explicit example of how risk improvements for estimating
$\mu$ are obtained by shrinking $X$ toward 0 by the adaptive
multiplicative factor $ (1- \frac{p-2}{\|x\|^2} )$.
Similarly,\vadjust{\goodbreak} under unimodal priors, posterior mean Bayes rules $\hat{\mu
}_\pi(x) = E_\pi( \mu\C x)$ shrink $x$ toward the center of~$\pi
(\mu)$, the mean of $\pi(\mu)$ when it exists. (Section~\ref
{sec:multshrink} will describe how multimodal priors yield multiple
shrinkage estimators.) As we saw earlier, $x$ here plays the role both
of $\hat\mu_{\mathrm{MLE}}(x) = x$ and of the formal Bayes estimator $\hat\mu
_{U}(x) = x$.

The representation (\ref{eq:mform}) reveals how $\hat{p}_\pi(y \C
x)$ analo\-gously ``shrinks'' the formal Bayes estimator $\hat{p}_U(y \C
x)$, but not $\hat{p}_{\mathrm{MLE}} \ne\hat{p}_U$, by an adaptive
multiplicative factor
%
\begin{equation}\label{eq:bfactor}
b_\pi(x,y) = \frac{m_\pi(w; v_w)} {m_\pi(x ; v_x)}.
\end{equation}
However, because $\hat{p}_\pi(y \C x)$ must be a proper probabi\-lity
distribution (whenever $m_\pi$ is always finite), it cannot be the
case that $b_\pi(x,y)\,{<}\,1$ for all $y$ at any~$x$. Thus, ``shrinkage''
here really refers to a~reconcentration of the probability distribution
of~$\hat{p}_U(y \C x)$. Furthermore, since the mean of~$\hat{p}_\pi
(y \C x)$ is~$E_\pi( \mu\C x)$, this reconcentration, under unimodal
priors, is toward the center of $\pi(\mu)$, as in the mean estimation case.

Consider, for example, what happens under $\pi_H$ which is symmetric
and unimodal about 0. Figure~\ref{fig1} illustrates how this shrinkage occurs
for $p_H$ for various values of $x$ when $p = 5$. Figure~\ref{fig1} plots $\hat
{p}_U(y \C x)$ and $\hat{p}_H(y \C x)$ as functions of $y =
(y_1,y_2,0,0,0)'$ when $v_x=1$ and $v_y=0.2$. Note first that $\hat
{p}_U(y \C x)$ is always the same symmetric shape centered at $x$. When
$x = (2,0,0,0,0)'$, shrinkage occurs by pushing the concentration of
$\hat{p}_H(y \C x)$ = $b_H(x,y) \hat{p}_U(y \C x)$ toward 0. As $x$
moves further from $(0, 0, 0, 0, 0)'$ to $(3,0,0,0,0)'$ and
$(4,0,0,0,0)'$ this shrinkage diminishes as $\hat{p}_H(y \C x)$
becomes more and more similar to $\hat{p}_U(y \C x)$.

\begin{figure}

\includegraphics{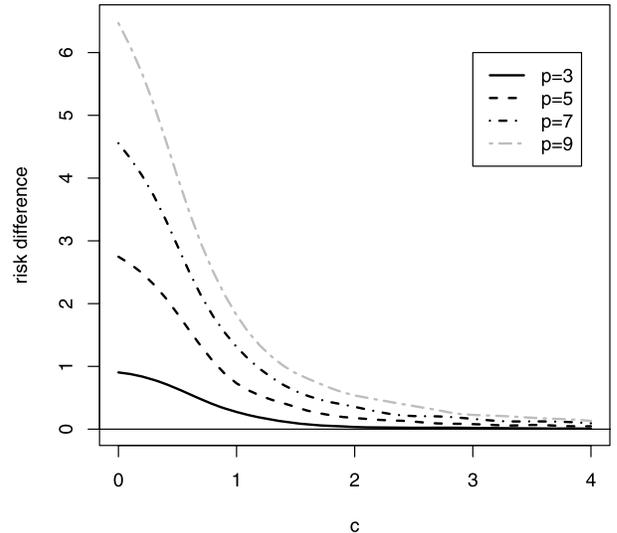}

\caption{The risk difference between $\hp_U$ and $\hp_H$ when $\mu = (c, \ldots,
c)'$, $v_x = 1$, $v_y = 0.2$.}
\label{fig2}\end{figure}

As in the problem of mean estimation, the shrinkage by $\hat{p}_H$
manifests itself in risk reduction over~$\hat{p}_U$. To illustrate
this, Figure~\ref{fig2} displays the risk difference $[R_{\mathrm{KL}}(\mu, \hat{p}_U)
- R_{\mathrm{KL}}(\mu, \hat{p}_H)]$ at $\mu= (c,\ldots, c)'$, $0 \le c \le4$
when $v_x=1$ and $v_y=0.2$ for dimensions $p=3, 5, 7, 9$. Paralleling
the risk reduction offered by~$\hat{\mu}_H$ in the mean estimation
problem, the largest risk reduction offered by $\hat{p}_H$ occurs
close to $\mu=0$ and decreases rapidly to 0 as $\|\mu\|$ increases.
[$R_{\mathrm{KL}}(\mu,\break \hat{p}_U)$ is constant as a function of $\mu$.] At
the same time, the risk reduction by $\hat{p}_H$ is larger for larger
$p$ at each fixed $\|\mu\|$.

\section{Many Possible Shrinkage Targets}\label{sec5}\label{sec:targets}

By a simple shift of coordinates, the modified James--Stein estimator,
%
\begin{equation}\label{JSb}
\hat\mu_{\mathrm{JS}}^b(x) = b +  \biggl(1- \frac{p-2}{\|x -b\|^2} \biggr) (x-b),
\end{equation}
remains minimax, but now shrinks $x$ toward $b \in R^p$ where its risk
function is smallest. Similarly, minimax Bayes shrinkage estimators of
a mean or of a~predictive density can be shifted to shrink toward~$b$,
by recentering the prior $\pi(\mu)$ to $\pi^b(\mu) = \pi(\mu-b)$.
These shifted estimators are easily obtained by inserting the%
corresponding translated mar\-ginal
%
\begin{equation}
m_\pi^b(z;v) = m_\pi(z-b;v)
\end{equation}
into (\ref{keyrep1}) to obtain
%
\begin{equation}\label{keyrepb}
\hat\mu_\pi^b(x) = E_\pi^b(\mu\C x) = x + \nabla\log m_\pi^b(x ; 1),
\end{equation}
and into (\ref{eq:mform}) to obtain
%
\begin{equation}
\hat{p}_\pi^b(y \C x) = \frac{m_\pi^b (w; v_w)} {m_\pi^b(x ;
v_x)}  \hat{p}_U(y \C x).
\end{equation}
Recentered unimodal priors such as $\pi_H^b$ and $\pi_a^b$ yield
estimators that now shrink $x$ and $\hat{p}_U(y \C x)$ toward $b$
rather than toward 0. Since the superharmonic properties of $m_\pi$
are inherited by $m_\pi^b$, the minimaxity of such estimators will be
preserved.

In his discussion of Stein
(\citeyear{Ste62N1}), Lindley (\citeyear{Ste62N2}) noted that the
James--Stein estimator\vadjust{\goodbreak} could be modified to shrink toward $(\bar
{x},\ldots,\bar{x})' \in R^p$ ($\bar{x}$ is the mean of the components
of $x$), by replacing $b$ and \mbox{$(p-2)$} in~(\ref{JSb}) by $(\bar
{x},\ldots,\bar{x})'$ and $(p-3)$, respectively. The resulting estimator
remains minimax as long as \mbox{$p \ge4$} and offers smallest risk when $\mu
$ is close to the subspace of $\mu$ with identical coordinates, the
subspace spanned by the vector $1_p = (1,\ldots,1)'$. Note that $(\bar
{x},\ldots,\bar{x})'$ is the projection of $x$ into this subspace.

More generally, minimax Bayes shrinkage estimators of a mean or of a
predictive density can be similarly modified to obtain shrinkage toward
any (possibly affine) subspace $B \subset R^p$, whenever they
correspond to spherically symmetric priors. Such priors, which include
$\pi_H$ and $\pi_a$, are functions of $\mu$ only through $\|\mu\|$.
Such a modification is obtained by recentering the prior $\pi(\mu)$
around $B$ via
%
\begin{equation}\label{eq:pHB}
\pi^B(\mu) = \pi(\mu-P_B\mu),
\end{equation}
where $P_B\mu= \mbox{argmin}_{b \in B} \|\mu-b\|$ is the projection
of~$\mu$ onto $B$.
Effectively, $\pi^B(\mu)$ puts a uniform prior on~$P_B\mu$ and
applies a suitably modified version of~$\pi$ to $(\mu-P_B\mu)$. Note
that the dimension of $(\mu-P_B\mu)$, namely $(p - \dim(B))$, must
be taken into account when determining the appropriate modification
for~$\pi$. For example, recentering the harmonic prior $\pi_H(\mu) = \|
\mu\|^{-(p-2)}$ around the subspace spanned by $1_p$ yields
%
\begin{equation}\label{eq:pH1}
\pi_H^B(\mu) = \|\mu- \bar\mu1_p\|^{-(p-3)},
\end{equation}
where $\bar\mu= \mu'1_p/p$. Here, the uniform prior is put on
$P_B\mu= \bar\mu1_p$, and the harmonic prior in dimension $(p - \dim
(B)) = (p-1)$ (which is different from the harmonic prior in $R^p$) is
put on $(\mu- \bar\mu1_p)$, the orthogonal complement of $B$.

The marginal $m_\pi^B$ corresponding to the recente\-red~$\pi^B$ in
(\ref{eq:pHB}) can be directly obtained by recentering the spherically
symmetric marginal $m_\pi$ corresponding to $\pi$, that is,
%
\begin{equation}\label{eq:mB}
m_\pi^B(z;v) = m_\pi(z-P_Bz ;v),
\end{equation}
where $P_Bz$ is the projection of $z$ onto $B$. Analogously to $\pi
^B(\mu)$, $m_\pi^B(z;v)$ is uniform on $P_Bz$ and applies a suitably
modified version of $m_\pi$ to $(z-P_Bz)$. Here, too, the dimension of
$(z-P_Bz)$, namely $(p - \dim(B))$, must be taken into account when
determining the appropriate modification for $m_\pi$. For example,
recentering the marginal $m_\pi$ around the subspace spanned by $1_p$
would entail replacing $\|z\|$ by $\|z - \bar z 1_p\|$, where $\bar z =
z'1_p/p$, and appropriately modifying $m_\pi$\vspace*{2pt} to apply to $R^{p-1}$.

Applying the recentering (\ref{eq:pHB}) to priors such as~$\pi_H$ and
$\pi_a$, which are unimodal around 0, yields priors~$\pi_H^B$ and
$\pi_a^B$ and hence marginals $m_H^B$ and $m_a^B$, which are
unimodal
around $B$. Such recentered mar\-ginals yield mean estimators
%
\begin{equation}\label{keyrepB}
\hat\mu_\pi^B(x) = E_\pi^B(\mu\C x) = x + \nabla\log m_\pi^B(x ; 1),\vadjust{\goodbreak}
\end{equation}
and predictive density estimators
%
\begin{equation}\label{eq:Bform}
\hat{p}_\pi^B(y \C x) = \frac{m_\pi^B (w; v_w)} {m_\pi^B(x ;
v_x)}  \hat{p}_U(y \C x),
\end{equation}
that now shrink $x$ and $\hat{p}_U(y \C x)$ toward $B$ rather than
toward 0.
Shrinkage will be largest when $x \in B$, and will diminish as $x$
moves away from $B$. These estimators offer smallest risk when $\mu\in
B$, but do not improve in any important way over $x$ and $ \hat{p}_U(y
\C x)$ when $\mu$ is far from $B$.

A superharmonic $m_\pi$ will lead to a superharmo\-nic~$m_\pi^B$ as
long as $(p - \dim(B))$ is large enough. For example, the recentered
marginal $m_H^B$ will be superharmonic only when $(p - \dim(B)) \ge
3$. In such cases, the minimaxity of both $\hat\mu_\pi^B$
and $\hat{p}_\pi^B$ will be preserved.

\section{Where to Shrink?}\label{sec6}\label{sec:multshrink}

Stein's discovery of the existence of minimax shrin\-kage estimators such
as $\hat\mu_{\mathrm{JS}}^b(x)$ in (\ref{JSb}) demonstrated that costless
improvements over the minimax $\hat\mu_{\mathrm{MLE}}$ were available near any
target preselected by the statistician. As Stein
(\citeyear{Ste62N1}) put it when
referring to the use of such an estimator to center a confidence
region, the target ``should be chosen\ldots\ as one's best guess'' of $\mu
$. That frequentist considerations had demonstrated the folly of
ignoring subjective input was quite a shock to the perceived
``objectivity'' of the frequentist perspective.

Although the advent of minimax shrin\-kage estimators of the form $\hat
\mu_\pi^B$ in (\ref{keyrepB}) and $\hat{p}_\pi^B$ in (\ref
{eq:Bform}) opened up the possibility of small risk near any
preselected (affine) subspace $B \subset R^p$ (this includes the
possibility that $B$ is a single point), it also opened up a~challenging new problem, how to best choose such a $B$. From the vast
number of possible choices, the goal was to choose $B$ close to the
unknown $\mu$, otherwise risk reduction would be negligible. To add to
the difficulties, low-dimensional $B$, which offered the greatest risk
reduction, were also the most difficult to get close to $\mu$.

When faced with a number of potentially good target choices, say
$B_1,\ldots,B_N$, rather than choose one of them and proceed with
$\hat\mu_\pi^B$ or $\hat{p}_\pi^B$, an attractive alternative is
to use a minimax multiple shrinkage estimator; see George (\citeyear{Geo86N1,Geo86N2,Geo86N3}).
Such estimators incorporate all the potential targets by combining
them
into an adaptive convex combination of
$\hat\mu_\pi^{B_1},\ldots,\hat\mu_\pi^{B_N}$ for mean estimation,
and of $\hat{p}_\pi^{B_1},\ldots,\hat{p}_\pi^{B_N}$ for predictive
density estimation. By adaptively shrinking toward the more promising
targets, the region of potential risk reduction is vastly enlarged
while at the same time retaining the safety of minimaxity.

The construction of these minimax multiple shrinkage estimators
proceeds as follows, again making fundamental use of the Bayesian
formulation. For a~spherically symmetric prior $\pi(\mu)$, a set of
subspaces $B_1,\ldots,B_N$ of $R^p$, and a set of nonnegative weights
$w_1,\ldots,w_N$ such that $\sum_1^N w_i = 1$, consider the mixture prior
%
\begin{equation}\label{eq:priorstar}
\pi_*(\mu) = \sum_{i=1}^N w_i  \pi^{B_i}(\mu),
\end{equation}
where each $\pi^{B_i}$ is a recentered prior as in (\ref{eq:pHB}). To
simplify notation, we consider the case where\break each~$\pi^{B_i}$ is a
recentering of the same $\pi$, although in principle such a
construction could be applied with different priors. The marginal $m_*$
corresponding to the mixture prior $\pi_*$ in (\ref{eq:priorstar}) is
then simply
%
\begin{equation}\label{eq:mstar}
m_*(z; v) = \sum_1^N w_i   m_\pi^{B_i}(z;v),
\end{equation}
where $m_\pi^{B_i}$ are the recentered marginals\vspace*{1pt} corresponding to the
$\pi^{B_i}$ as given by (\ref{eq:mB}).

Applying Brown's representation $\hat\mu_\pi= x + \break \nabla\log m_\pi
(x ;1)$ from (\ref{keyrep1}) with $m_*$ in (\ref{eq:mstar})
immediately yields the multiple shrinkage estimator of $\mu$,
%
\begin{equation}\label{pstar1}
\hat{\mu}_*(x) = \sum_{i = 1}^N  p(B_i \C x)  \hat{\mu}_\pi^{B_i}(x),
\end{equation}
where
%
\begin{equation}\label{pprob1}
p(B_i \C x) = \frac{w_i m_\pi^{B_i}(x ; 1)} {\sum_{i = 1}^N w_i
m_\pi^{B_i}(x ; 1)}.
\end{equation}

Similarly, applying the representation $\hat{p}_\pi(y \C x) =\frac
{m_\pi(w; v_w)} {m_\pi(x ; v_x)}  \hat{p}_U(y \C x)$ from (\ref
{eq:mform}) with $m_*$\vspace*{2pt} immediately\break yields the multiple shrinkage
estimator of $p(y \C\mu)$,
%
\begin{equation}\label{pstar2}
\hat{p}_*(y \C x) = \sum_{i = 1}^N  p(B_i \C x)  \hat{p}_\pi
^{B_i}(y \C x),
\end{equation}
where
%
\begin{equation}\label{pprob2}
p(B_i \C x) = \frac{w_i m_\pi^{B_i}(x ; v_x)} {\sum_{i = 1}^N w_i
m_\pi^{B_i}(x ; v_x)}.
\end{equation}

The forms (\ref{pstar1}) and (\ref{pstar2}) reveal $\hat{\mu}_*$
and $\hat{p}_*$ to be adaptive convex combination of the individual
posterior mean estimators $\hat{\mu}_\pi^{B_i}$ and $\hat{p}_\pi
^{B_i}$, respectively. The adaptive weights $p(B_i \C x)$ in (\ref
{pprob1}) and (\ref{pprob2}) are the posterior probabilities that $\mu
$ is contained in each of the $B_i$, effectively putting increased
weight on those individual estimators which are shrinking most. Note
that the uniform prior estimates~$\hat{\mu}_U$\break and~$\hat{p}_U$ are
here doubly shrunk by $\hat{\mu}_*$ and $\hat{p}_*(y \C x)$; in
addition to the individual estimator shrinkage they are further shrunk
by the posterior probability $\hat{p}(B_i \C x)$.

The key to obtaining $\hat{\mu}_*$ and $\hat{p}_*(y \C x)$ which are
minimax is simply to use priors which yield superharmonic $m_\pi
^{B_1},\ldots, m_\pi^{B_N}$. If such is the case, then trivially from
(\ref{eq:mstar})
%
\begin{equation}
\nabla^2 m_* = \sum_1^N w_i \nabla^2 m_\pi^{B_i} \le0,
\end{equation}
so that $m_*$ will be superharmonic, and the minimaxity of $\hat{\mu
}_*$ and $\hat{p}_*(y \C x)$ will follow immediately. Note that
marginals whose square root is superharmonic will not be adequate, as
this argument will fail.

\begin{figure}[b]
\includegraphics{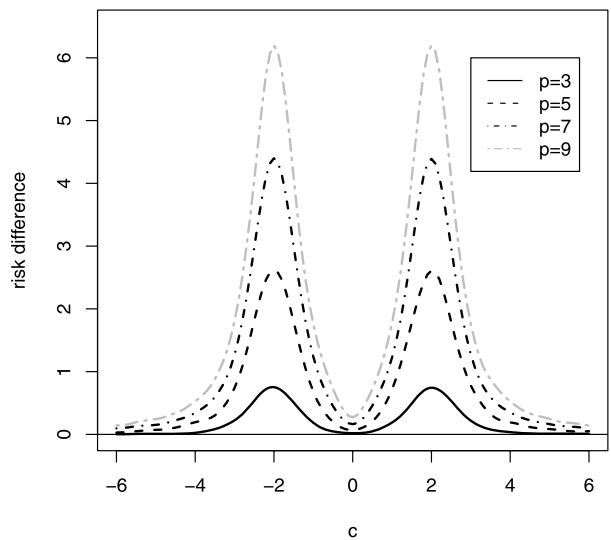}

\caption{The risk difference between $\hp_U$ and multiple shrinkage $\hp_{H^*}$
when $\mu = (c, \ldots, c)'$, $v_x = 1$, $v_y = 0.2$, $b_1 = (2,\ldots
,2)'$, $b_2 = (-2,\dots,-2)'$, and $w_1 = w_2 = 0.5$.}
\label{fig3}\end{figure}

The adaptive shrinkage behavior of $\hat{\mu}_*$ and $\hat{p}_*$
ma\-nifests itself as substantial risk reduction whenever $\mu$ is near
any of $B_1,\ldots, B_N$. Let us illustrate how that happens for the
predictive density estimator $\hat{p}_{H^*}$, the multiple shrinkage
version of $\hat{p}_{H}$. Figure~\ref{fig3} illustrates the risk reduction
$[R_{\mathrm{KL}}(\mu, \hat{p}_U) - R_{\mathrm{KL}}(\mu, \hat{p}_{H^*})]$ at various
$\mu= (c,\ldots, c)'$ obtained by $\hat{p}_{H^*}$ which adaptively
shrinks $\hat{p}_U(y \C x)$ toward the closer of the two points $b_1 =
(2,\ldots,2)'$ and $b_2 = (-2,\ldots,-2)'$ using equal weights $w_1 =
w_2 = 0.5$. As in Figure~\ref{fig2}, we considered the case $v_x=1, v_y=0.2$ for
$p=3, 5, 7, 9$. As the plot shows, maximum risk reduction occurs when
$\mu$ is close to $b_1$ or $b_2$, and goes to 0 as $\mu$ moves away
from either of these points. At the same time, for each fixed $\|\mu\|
$, risk reduction by $\hat{p}_{H^*}$ is larger for larger $p$. It is
impressive that the size of the risk reduction offered by $\hat
{p}_{H^*}$ is nearly the same as each of its single target
counterparts. The cost of multiple shrinkage enhancement seems
negligible, especially compared to the benefits.

\section{Empirical Bayes Constructions}\label{sec7}

Beyond their attractive risk properties, the James--Stein estimator
$\hat\mu_{\mathrm{JS}}$ and its positive-part counterpart $\hat\mu_{\mathrm{JS}+}$
are especially appealing because of their simple closed forms which are
easy to compute. As shown by Xu
and Zhou (\citeyear{XuZho11}), similarly appealing
simple closed-form predictive density shrinkage estimators can be
obtained by the same empirical Bayes considerations that motivate $\hat
\mu_{\mathrm{JS}}$ and $\hat\mu_{\mathrm{JS}+}$.

The empirical Bayes motivation of $\hat\mu_{\mathrm{JS}}$, alluded to in
Section \ref{sec1}, simply entails replacing $1/(1+\nu)$ in (\ref
{conjugate-coef}) by $(p-2)/\|x\|^2$, its unbiased estimate under the
marginal distribution of $X \C\mu\sim N_p(\mu, I)$ when $\mu\sim
N_p(0, \nu I)$. The positive-part $\hat\mu_{\mathrm{JS}+}$ is obtained by
using the truncated estimate $(p-2)/\max\{1,\|x\|^2\}$ which avoids an
implicitly negative estimate of the prior variance $\nu$.

Proceeding analogously, Xu and Zhou considered the Bayesian predictive
density estimate,
%
\begin{eqnarray}
&&\hspace*{-4pt}  \hat p_{\nu}(y \C x) \sim N_p \biggl( \biggl(1-\frac{v_x}{v_x + \nu
} \biggr)x, \nonumber
\\[-8pt]
\\[-8pt]
 &&\hspace*{65pt}\frac{v_x}{v_x + \nu}v_y +  \biggl(1-\frac{v_x}{v_x + \nu
} \biggr)(v_x+v_y)  \biggr),
\nonumber
\end{eqnarray}
when $X \C\mu\sim N_p(\mu, v_x I)$ and $Y \C\mu\sim N_p(\mu, v_y
I)$ are independent, and
$\mu\sim N_p(0,\nu I)$. Replacing $v_x/(v_x +\nu)$ by its truncated
unbiased estimate $(p-2)v_x/\allowbreak\max\{v_x,  \|x\|^2\}$ under the marginal
distribution of $X$, they obtained the empirical Bayes predictive
density estimate
%
\begin{eqnarray} \label{eq:positive_JS_hp}
&&\hat p_{p-2}(y \C x) \sim N_p  \biggl( \biggl(1-\frac{(p-2) v_x}{\|x\|
^2} \biggr)_+ x;\nonumber
\\[-8pt]
\\[-8pt]   &&\hphantom{\hat p_{p-2}(y \C x) \sim N_p  \biggl(}
v_y +  \biggl(1-\frac{(p-2)v_x}{\|x\|^2}
\biggr)_+ v_x  \biggr)
\nonumber
\end{eqnarray}
where $(\cdot)_+ = \max\{0,\cdot\}$, an appealing simple closed
form. Centered at $\hat\mu_{\mathrm{JS}+}$, $\hat p_{p-2}$ converges to the
best invariant procedure $\hat p_U \sim N(x, v_x+v_y)$ as $\|x\|^2
\rightarrow\infty$, and\vadjust{\goodbreak
} converges to $N(0, v_y)$ as $\|x\|^2
\rightarrow0$. Thus, $\hat p_{p-2}$ can be viewed as a shrinkage
predictive density estimator that ``pulls'' $\hat p_U$ toward $0$, its
shrinkage adaptively determined by the data.

To assess the KL risk properties of such empirical Bayes estimators, Xu
and Zhou considered the class of estimators $\hat p_k$ of the form
(\ref{eq:positive_JS_hp}) with $(p-2)$ replaced by a constant $k$, a
class of simple normal forms centered at shrinkage estimators of $\mu$
with data-dependent variances to incorporate estimation uncertainty.
For this class, they provided general sufficient conditions on $k$ and
the dimension $p$ for~$\hat p_k$ to dominate the best invariant
predictive density~$\hat p_U$ and thus be minimax. Going further, they
also established an ``oracle'' inequality which suggests that the
empirical Bayes predictive density estimator is asymptotically minimax
in infinite-dimensional parameter spaces and can potentially be used to
construct adaptive minimax estimators. It appears that these minimax
empirical Bayes predictive densities may play the same role as the
James--Stein estimator in such problems.

It may be of interest to note that a particular pseudo-marginal
empirical Bayes construction that works fine for the mean estimation
problem appears not to work for the predictive density estimation
problem. For instance, the positive-part James--Stein estimator $\hat
\mu_{\mathrm{JS}+}$ can be expressed as $\hat\mu_{\mathrm{JS}+} = x +\break \nabla\log
m_{\mathrm{JS}+}(x ; 1)$, where $m_{\mathrm{JS}+}(x; v)$ is the function
\begin{eqnarray*}\label{eq:mS2}
&&m_{\mathrm{JS}+}(x; v)\\
 & & \quad
= \cases{ k_p   \|x\|^{-(p-2)}     \quad  \mbox{if    $\|x\|^2/v \ge
(p-2)$}, \vspace*{3pt}\cr  v^{-(p-2)/2}  \exp\{-\|x\|^2/2v \} \vspace*{1pt} \cr   \hspace*{68.5pt} \mbox{if   $\|x\|^2/v <
(p-2)$},}
\end{eqnarray*}
with $k_p = (e /(p-2))^{-(p-2)/2}$ (see Stein,
\citeyear{Ste74}). We refer to
$m(z;v)$ as a pseudo-marginal because it is not a bona fide marginal
obtained by a real prior. Nonetheless, it plays the formal role of a
marginal in the mean estimation problem, and can be used to generate
further innovations such as minimax multiple shrinkage James--Stein
estimators (see George, \citeyear{Geo86N1,Geo86N2,Geo86N3}).

Proceeding by analogy, it would seem that $m(z;v)$ could be inserted
into the representation (\ref{eq:mform}) from Lemma \ref{thm:pform}
to obtain similar results under KL loss.
Unfortunately, this does not yield a suitable minimax predictive
estimator because $\hat{p}_{\mathrm{JS}+}(y \C x)$ is not a~proper probability
distribution. Indeed, $\int\hat{p}_{\mathrm{JS}+}(y \C\break x)\,dy \ne1$ and varies
with $x$. What has gone wrong? Because they do not correspond to real
priors, such pseudo-marginals are ultimately at odds with the
probabilistic coherence of a valid Bayesian approach. In contrast to
the mean estimation framework, the predictive density estimation
framework apparently requires stronger fidelity to the Bayesian paradigm.

\section{Predictive Density Estimation for Classical Regression}\label{sec8}

Moving into the multiple regression setting, Stein
(\citeyear{Ste60}) considered
the estimation of a $p$-dimensional coefficient vector under suitably
rescaled quadratic loss. He there established the minimaxity of the
maximum likelihood estimators, and then proved its inadmissibility when
$p\ge3$, by demonstrating the existence of a dominating shrinkage estimator.

In a similar vein, as one might expect, the theory of predictive
density estimation presented in Sections \ref{sec:pred-emerges} and
\ref{sec:theory} can also be extended to the multiple regression framework.
We here describe the main ideas of the development of this extension
which appeared in George
and Xu (\citeyear{GeoXu08}).
Similar results, developed independently from a slightly different
perspective, appeared at the same time in Kobayashi
and Komaki (\citeyear{KobKom08}).

Consider the canonical normal linear regression setup:
%
\begin{equation}
 \qquad X \C\beta\sim N_m(A \beta, \sigma^2 I), \quad Y \C\beta\sim N_n(B
\beta, \sigma^2 I),
\end{equation}
where $A$ is a full rank, fixed $m \times p$, $B$ is a fixed $n \times
p$ matrix, and $\beta$ is a common $p\times1$ unknown regression
coefficient. The error variance $\sigma^2$ is assumed to be known, and
set to be $1$ without loss of generality. The problem is to find an
estimator of $\hat p(y \C x)$ of the predictive density $p(y \C \beta
)$, evaluating its performance by KL risk
%
\begin{equation}
R_{\mathrm{KL}}(\beta, \hat p) = \int p(x \C \beta) L(\beta, \hat p(\cdot\C
x))\,dx,
\end{equation}
where $L(\beta, \hat p( \cdot\C x))$ is the KL loss between the
density $p(y \C\beta)$ and its estimator $\hat p(y \C x).$

The story begins with the result, analogous to Aitchison's
(\citeyear{Ait75}) for the normal mean problem, that the plug-in
estimator $p(y \C\hat\beta_x)$, where $\hat{\beta}_x$ is the least
squares estimate of $\beta$ based on $x$, is dominated under KL risk
by the posterior mean of $p(y \C\beta)$, the Bayes rule under the
uniform prior
%
\begin{eqnarray}
\hat{p}_U(y \C x) &=& \frac{1}{(2 \pi)^{\fraca{n}{2}}} \frac{|A'A +
B'B|^{-\fraca{1}{2}}}{|A' A|^{-\fraca{1}{2}}} \nonumber
\\[-8pt]
\\[-8pt]
&&{}\times\exp \biggl\{ - \frac
{{\mathit{RSS}}_{x,y} - {\mathit{RSS}}_{x}}{2}  \biggr\}.
\nonumber
\end{eqnarray}
Here, too, $\hat{p}_U$ is minimax (Liang,
\citeyear{Lia02}; Liang
and Barron, \citeyear{LiaBar04}) and plays the straw-man role of\vadjust{\goodbreak} the estimator to beat. The
challenge was to determine which priors $\pi$ would lead to Bayes
rules which dominated $\hat{p}_U$, and hence would be minimax.
Analogously to the representation (\ref{eq:mform}) in Lemma \ref
{thm:pform} for the normal mean problem, the following representation
for a Bayes rule $\hat{p}_\pi(y \C x)$ here, was the key to meeting
this challenge.

%
\begin{lemma}\label{thm:pform:reg} The Bayes rule $\hat{p}_\pi(y \C
x) = \int p(y \C\beta)\* \pi(\beta)\,d\beta$ can be expressed as
%
\begin{equation} \label{eq:mform:reg}
\hat{p}_\pi(y \C x) =\frac{m_\pi(\hat{\beta}_{x,y}; \Sigma_C)}
{m_\pi(\hat{\beta}_x ; \Sigma_A)}  \hat{p}_U(y \C x),
\end{equation}
where $\Sigma_A\,{=}\,(A' A)^{-1},$ $C\,{=}\,A'A\,{+}\,B' B,$ $\Sigma_C\,{=}\,(C'C)^{-1}$,
$\hat{\beta}_x$ is the least squares estimates of $\beta$
based on $x$, and~$\hat{\beta}_{x,y}$ based on $x$ and $y$, and
$m_\pi(z ; \Sigma)$ is the marginal distribution of $Z | \beta\sim
N_p(\beta, \Sigma)$ under $\pi(\beta)$.
\end{lemma}

The representation (\ref{eq:mform:reg}) leads immediately to the
following analogue of (\ref{eq:uber3}) for the KL risk difference
between $\hat{p}_U(y \C x)$ and $\hat{p}_\pi(y \C x)$:
%
\begin{eqnarray}
\label{eq:uber3:reg}
&&R_{\mathrm{KL}}(\beta, \hat{p}_U) - R_{\mathrm{KL}}(\beta, \hat{p}_\pi)
\nonumber\\
&& \quad = E_{\beta, \Sigma_C} \log m_\pi( \hat{\beta}_{x,y}; \Sigma_C)
\\&& \qquad {} -
E_{\beta, \Sigma_A} \log m_\pi( \hat{\beta}_{x}; \Sigma_A).
\nonumber
\end{eqnarray}
The challenge thus became that of finding conditions on $m_\pi$ to
make this difference positive, a challenge made more difficult than the
previous one for (\ref{eq:uber3}) because of the complexity of $\Sigma
_A$ and $\Sigma_C$. Fortunately this could be resolved by rotating the
problem as follows to obtain diagonal forms.
Since $\Sigma_A$ and $\Sigma_C$ are both symmetric and positive
definite, there exists a full rank $p\times p$ matrix $W$, such that
%
\begin{eqnarray}\label{eq:W:reg}
\Sigma_A &=& W W', \quad\Sigma_C = W D W', \nonumber
\\[-8pt]
\\[-8pt] D&=& \operatorname{diag}(d_1,\ldots, d_p).
\nonumber
\end{eqnarray}
Because $\Sigma_C = (\Sigma_A^{-1} + B'B)^{-1}$ where $B' B$ is
nonnegative definite, it follows that $d_i \in(0,1]$ for all $1 \le i
\le p$ with at least one $d_i<1$. Thus, the parameters for the rotated
problem become
%
\begin{eqnarray}
\mu&=& W^{-1} \beta, \quad\hat{\mu}_x = W^{-1} \hat{\beta}_x \sim
N_p(\mu, I), \nonumber
\\[-8pt]
\\[-8pt]\hat{\mu}_{x,y} &=& W^{-1} \hat{\beta}_{x,y} \sim
N_p(\mu, D).
\nonumber
\end{eqnarray}
Letting $V_w = w I + (1-w) D$ for $w \in[0,1]$, the risk difference
(\ref{eq:uber3:reg}) could be reexpressed as
%
\begin{eqnarray}
&&R_{\mathrm{KL}}(\beta, \hat{p}_U) - R_{\mathrm{KL}}(\beta,
\hat{p}_\pi)\nonumber\\
&& \quad = E_{\mu, D} \log m_{\pi_W} ( \hat{\mu}_{x,y}; D) \nonumber
\\[-8pt]
\\[-8pt]&&{} \qquad - E_{\mu, I}
\log m_{\pi_W} ( \hat{\mu}_{x}; I) \nonumber\\
&& \quad = h_{\mu}(V_0) - h_{\mu}(V_1),
\nonumber
\end{eqnarray}
where $h_{\mu}(V_w) = E_{\mu, V_w} \log m_{\pi_W} ( Z; V_w)$ and
$\pi_W (\mu) = \pi(W \mu)$. The minimaxity of $\hat{p}_\pi$ would
now follow from conditions on $m_{\pi}$ such that $(\partial/
\partial w) h_{\mu}(w) < 0$ for all $\mu$ and $w \in[0,1].$ The
following substantial generalizations of Theorem~\ref{theo1} and
Corollary~\ref{cor1}
provide exactly those conditions.

%
\begin{theorem} Suppose $m_{\pi}(z; W W')$ is finite for all $z$ with
the invertible matrix $W$ defined as in (\ref{eq:W:reg}). Let
$H(f(z_1, \ldots, z_p))$ be the Hessian matrix of $f$.
\begin{longlist}[(ii)]
\item[(i)] If $\operatorname{trace}  \{H(m_{\pi}(z; W V_w W')) [ \Sigma_A - \Sigma
_C]  \} \le0$\break for all $w \in[0,1]$, then $\hat{p}_{\pi}(y \C
x)$ is minimax.
\item[(ii)] If $\operatorname{trace}  \{H( \sqrt{m_{\pi}(z; W V_w W')}) [ \Sigma_A -
\Sigma_C]  \} \le0$\break for all $w \in[0,1]$, then $\hat{p}_{\pi
}(y \C x)$ is minimax.
\end{longlist}
\end{theorem}
%

%
\begin{coro} Suppose $m_{\pi}(z; W W')$ is finite\break for all $z$. Then
$\hat{p}_{\pi}(y \C x)$ is minimax if
\[
\operatorname{trace}  \{H(\pi(\beta)) [ \Sigma_A - \Sigma_C]  \} \le0   \quad
a.e.
\]
\end{coro}

As a consequence of Corollary 2, the scaled harmonic prior $\pi
_{H}(\beta| W) \propto\| W^{-1} \beta\|^{p-2}$ can be shown to yield
minimax predictive density estimators for the regression setting.

Going further, George
and Xu (\citeyear{GeoXu08}) went on to show that the minimax
Bayes estimators here can be modified to shrink toward different points
and subspaces as in Section~\ref{sec5}, and that the minimax multiple shrinkage
constructions of Section~\ref{sec6} apply as well. In particular, they obtained
minimax multiple shrinkage estimators that naturally accommodate
variable selection uncertainty.

\section{Predictive Density Estimation for Nonparametric Regression}\label{sec9}

Moving in another direction, Xu
and Liang (\citeyear{XuLia10}) considered predictive
density estimation in the context of modern nonparametric regression, a
context in which the James--Stein estimator has turned out to play an
important asymptotic minimaxity role; see Wasserman
(\citeyear{Was06}). Their
results pertain to the canonical setup for nonparametric regression:
%
\begin{eqnarray} \label{eq:X:nonpa}
Y(t_i) = f(t_i) + \varepsilon_i, \quad i = 1, \ldots, n,
\end{eqnarray}
where $f$ is an unknown smooth function in $\mathcal{L}^2[0, 1]$,
$t_i=i/n$, and $\varepsilon_i$'s are i.i.d. $N(0,1)$. A central
problem here is to estimate $f$ or various functionals of~$f$ based on
observing $Y=(Y(t_1),\ldots, Y(t_n))$. Transforming the problem with
an orthonormal basis, (\ref{eq:X:nonpa}) is equivalent to estimating
the $\theta_i$'s in
%
\begin{equation} \label{seq:model:nonpa}
 \qquad y_i = \theta_i + e_i, \quad e_i \sim N  \biggl(0, \frac{1}{n}  \biggr),
\quad i=1, \ldots, n,
\end{equation}
known as the Gaussian sequence model. The model above is different from
the ordinary multivariate normal model in two aspects: (1) the model
\mbox{dimension~$n$} is increasing with the sample size, and (2)~under
function space assumptions on $f$, the $\theta_i$'s lie in a~con\-strained space, for example, an ellipsoid $ \{ \sum_i a_i^2
\theta_i^2 \le C, a_i \to\infty \}$.

A large body of literature has been devoted to minimax estimation of
$f$ under $\mathcal{L}^2$ risk over
certain function spaces; see, for example, Johnstone (\citeyear{Joh}), Efromovich
(\citeyear{Efr99}), and the references therein. As opposed to the ordinary
multivariate normal mean problem, exact minimax analysis is difficult
for the Gaussian sequence model (\ref{seq:model:nonpa}) when a
constraint on the parameters is considered. This difficulty has been
overcome by first obtaining the minimax risk of a subclass of
estimators of a simple form, and then showing that the overall minimax
risk is asymptotically equivalent to the minimax risk of the subclass.
For example, an important result from Pinsker
(\citeyear{Pin80}) is that when the
parameter space is constrained to an ellipsoid, the nonlinear minimax
risk is asymptotically equivalent to the linear minimax risk, namely
the minimax risk of the subclass of linear estimators of the form $\hat
\theta_i = c_i x_i$.

For nonparametric regression, the following analogue between estimation
under $\mathcal{L}^2$ risk and predictive density estimation under KL
risk was established in Xu
and Liang (\citeyear{XuLia10}). The prediction problem for
nonparametric regression is formulated as follows. Let $\Y=(\Y(u_1),\ldots, \Y(u_m))$ be future
observations arising at a set of dense ($m \ge n$) and equally spaced
locations $\{u_j\}_{i=1}^m$.
Given $f$, the predictive density $p(\tilde{y}
\C f)$ is just a product of Gaussians. The problem is to find an
estimator $\hp(\tilde{y}\C y)$ of $p(\tilde{y}\C f)$, where
performance is measured by the averaged KL risk
%
\begin{equation} \label{risk:simultaneous:nonpa}
R(f, \hp) = \frac{1}{m} E_{Y, \Y| f} \log\frac{p(\Y\C
f)}{\hp(\Y\C Y)}.
\end{equation}
In this formulation, densities are estimated at the~$m$ locations
simultaneously by $\hp(\tilde{y}\C y)$. As it turned out, the KL risk
based on the simultaneous formula-\break tion~(\ref{risk:simultaneous:nonpa})
is the analog of the
$\mathcal{L}^2$ risk for estimation. Indeed, under the KL risk (\ref
{risk:simultaneous:nonpa}), the prediction problem for a nonparametric
regression model can be converted to the one for a Gaussian sequence model.

Based on this formulation of the problem, minimax analysis proceeds as
in the general framework for the minimax study of function estimation
used by, for example, Pinsker
(\citeyear{Pin80}) and Belitser
and~Le\-vit (\citeyear{BelLev95}, \citeyear{B96}). The linear estimators there, which play a central role in their
minimax analysis, take the same form as posterior means under normal
priors. Analogously, predictive density estimates under the same normal
priors turned out to play the corresponding role in the minimax
analysis for prediction. (The same family of Bayes rules arises from
the empirical Bayes approach in Section~\ref{sec7}.) Thus, Xu
and Liang (\citeyear{XuLia10})
were ultimately able to show that the overall minimax KL risk is
asymptotically equivalent to the minimax KL risk of this subclass of
Bayes rules, a direct analogue of Pinker's Theorem for predictive
density estimation in nonparametric regression.

\section{Discussion}\label{sec10}

Stein's
(\citeyear{Ste56}) discovery of the existence of shrinkage estimators that
uniformly dominate the minimax maximum likelihood estimator of the mean
of a multivariate normal distribution under quadratic risk when $p \ge
3$ was the beginning of a major research effort to develop improved
minimax shrinkage estimation. In subsequent papers Stein guided this
effort toward the Bayesian paradigm by providing explicit examples of
minimax empirical Bayes and fully Bayes rules. Making use of the
fundamental results of Brown
(\citeyear{Bro71}), he developed a general theory for
establishing minimaxity based on the superharmonic properties of the
marginal distributions induced by the priors.

The problem of predictive density estimation of a~multivariate normal
distribution under KL risk has more recently seen a series of
remarkably parallel developments. With a focus on Bayes rules catalyzed
by Aitchison (\citeyear{Ait75}), Komaki
(\citeyear{Kom01}) provided a fundamental breakthrough
by demonstrating that the harmonic prior Bayes rule dominated the best
invariant uniform prior Bayes rule. These results suggested the
existence of a theory for minimax estimation based on the superharmonic
properties of marginals, a theory that was then established in George,
Liang and Xu (\citeyear{GeoLiaXu06}). Further developments of new minimax shrinkage
predictive density estimators now abound, including, as described in
this article, multiple shrinkage estimators, empirical Bayes
estimators, normal linear model regression estimators, and
nonparametric regression estimators. Examples of promising further new
directions for predictive density estimation can be found in the work
of Komaki
(\citeyear{Kom04,Kom06,Kom09}) which included results for Poisson
distributions,\vadjust{\goodbreak} for general location-scale models and for Wishart
distributions, in the work of Ghosh,
Mergel and Datta (\citeyear{GhoMerDat08}) which
developed estimation under alternative divergence losses, and in the
work of Kato
(\citeyear{Kat09}) which established improved minimax predictive
domination for the multivariate normal distribution under KL risk when
both the mean and the variance are unknown. Minimax predictive density
estimation is now beginning to flourish.

\section*{Acknowledgments}
This work was supported by NSF Grants   DMS-07-32276 and DMS-09-07070.
The authors are grateful for the helpful comments and
clarifications of an anonymous referee.

%

\end{document}